\documentclass[a4paper,12pt]{article}

\usepackage[scale={.75,.8}]{geometry} % Wide page
\usepackage{pstricks}
\usepackage[dvips]{graphicx}
\usepackage{verbatim}
\usepackage[latin2]{inputenc}
\usepackage{epic}
\usepackage{latexsym}
\usepackage{epsf}
\usepackage{epsfig}
\usepackage{amssymb,amsmath}

\newcommand{\fref}[1]{fig.\ \ref{f.#1}} 
\newcommand{\eref}[1]{eq.\ (\ref{e.#1})} 
\newcommand{\erefn}[1]{ (\ref{e.#1})}
 
\newcommand{\aref}[1]{\ref{a.#1}}
\newcommand{\sref}[1]{Section \ref{s.#1}}
\newcommand{\cref}[1]{Chapter \ref{c.#1}}

\def\nn{\nonumber \\}  
\newcommand{\nl}{& \nonumber \\ &}

\def\beq{\begin{equation}} 
\def\eeq{\end{equation}} 
\newcommand{\ba}{\begin{array}}  
\newcommand{\ea}{\end{array}} 
\newcommand{\bea}{\begin{eqnarray}}  
\newcommand{\eea}{\end{eqnarray} }  
\newcommand{\bal}{\begin{align}}
\newcommand{\eal}{\end{align}}   
\def\bi{\begin{itemize}}  
\def\ei{\end{itemize}}  
\def\ben{\begin{enumerate}}  
\def\een{\end{enumerate}}  
\def\beq{\begin{equation}}  
\def\eeq{\end{equation}}  
\def\bc{\begin{center}}
\def\ec{\end{center}} 
 \def\bt{\begin{table}}
\def\et{\end{table}}  
 \def\btb{\begin{tabular}}
\def\etb{\end{tabular}}  
\newcommand{\bvec}{\left ( \ba{c}}
\newcommand{\evec}{\ea \right )}

\def\cl{{\mathcal L}}  
   
\def\cm{{\mathcal M}}  
\def\co{{\mathcal O}}

\def\gev{\, {\rm GeV}}
\def\tev{\, {\rm TeV}}

\def\mass2{mass${}^2$}

\newcommand{\mkk}{M_{\mathrm{KK}}}

\def\pa{\partial}

\newcommand{\tr}{\mathrm T \mathrm r}  

\def\rt{\sqrt{2}}

\def\ra{\rangle}
\def\la{\langle}

\newcommand{\ha}{{\hat a}}

\newcommand{\ti}{\tilde}  
\def\hc{{\rm h.c.}}

\def\ov{\overline}

\def\eps{\epsilon}

\begin{document}
  \begin{titlepage}
    \bigskip{} {\par\centering \textbf{
Pseudo-Goldstone Higgs Production via Gluon Fusion } \large \par}
    \bigskip{}
    
	    {\par\centering Adam Falkowski$^{1,2}$ \\ \par}
	    \bigskip{}
	    
		    {\par\centering
		      {
			{\small $^1$ CERN Theory Division, CH-1211 Geneva 23, Switzerland
			}\\
			{\small $^2$ Institute of Theoretical Physics, Warsaw University, 
			  \\ Ho\.za 69, 00-681 Warsaw, Poland}
			\par}
		    }
		    
		    \bigskip{}

		    \begin{abstract}
		      \noindent
The gluon-gluon-higgs amplitude is investigated in the context of 5D models of gauge-higgs unification.
A simple algorithm allows to include,  in a fully  analytical way, the contribution of the whole Kaluza-Klein tower of a 5D quark to the amplitude.     
This algorithm is applied to realistic models based on $SO(5)$ symmetry. 
Within the studied classes of models, the higgs production cross section is always suppressed.

		    \end{abstract}
		    %      }}}
		    %    \end{center}
\end{titlepage}

\newpage

\setcounter{footnote}{0}

\renewcommand{\theequation}{\arabic{section}.\arabic{equation}} 

%%%%%%%%%%%%%%%%%%%%%%%%%%%%%%%%%%%%%%%%%%%%%%%%%%%%%%%%%%%%%%%%%%%%%%
\section{Introduction}
\label{sec:intro}
%%%%%%%%%%%%%%%%%%%%%%%%%%%%%%%%%%%%%%%%%%%%%%%%%%%%%%%%%%%%%%%%%%%%%%

The higgs boson will soon be copiously produced  in the LHC, or so we believe.
Within the Standard Model (SM), the dominant production mechanism in hadron colliders is the  gluon fusion \cite{geglma,babis}.
This process is known to be particularly sensitive to new physics.  
New coloured particles at the TeV scale or below may significantly alter the SM predictions for the gluon fusion amplitude.
The hierarchy problem of the SM strongly suggests that such new physics states do exist and that they have sizable couplings to the higgs boson.  
The examples thoroughly studied in the literature include squarks in supersymmetry, vector-like quarks in little higgs and Kaluza-Klein (KK) quarks in higher dimensional scenarios.   
  
In this paper I investigate the one-loop gluon-gluon-higgs amplitude in 5D models of gauge-higgs unification \cite{ho}, which provide a dual realization of the pseudo-goldstone boson higgs scenario in 4D \cite{conopo,agcopo}.  
Higher dimensional models typically contain KK partners of the top quark whose Yukawa coupling to the higgs boson is of order $y_t m_t/\mkk$.
Moreover, the top quark Yukawa coupling is itself modified by terms of order $m_t^2/\mkk^2$.    
In 5D gauge-higgs unification, these couplings are further constrained by the fact that the radiatively generated higgs potential must be free of divergences.  
I will show that the intricate structure of the 5D models results in very robust predictions concerning the gluon fusion amplitude.  

In \sref{5d},  I describe a simple algorithm for computing the one-loop contribution of a full KK tower to the gluon fusion amplitude.
When all quarks in the tower are sufficiently heavy, $2 m_n > m_{\rm higgs}$,  the result can be expressed in terms of a low energy limit of the UV brane-to-brane propagators.
Thus, the task is reduced to finding mixed 4D momentum/5D position space propagators, 
which can be formally solved in an arbitrary 5D warped background.
The bottom line is that the gluon fusion amplitude can be calculated analytically, without resorting to numerical methods.     
 
In \sref{so5} this algorithm is applied to realistic models based on $SO(5)$ gauge symmetry in 5D \cite{agcopo,codapo}. 
The final result for the gluon fusion amplitude turns out to be surprisingly simple and does not depend on fine-grained details of the model.
The only continuous parameter  that enters the result is the global symmetry breaking scale $f$. 
%which, in pseudo-goldstone higgs scenarios, plays analogous role as the pion decay constant in low-energy QCD.  
The result depends also on the embedding of  the third generation quark sector into multiplets of the $SO(5)$ gauge symmetry.  
For the two embeddings studied in this paper, the gluon fusion amplitude is suppressed with respect to the SM prediction.  
This is a manifestation of the more general conjecture \cite{ian} that, in models solving the naturalness problem of the SM, the higgs production cross section is diminished.
The reason is that cancellation of quadratic divergences implies a particular structure of the Yukawa couplings of the higgs to the quarks, from which the suppression can be deduced  \cite{ian}.
The new element here is that the suppression can be quantified  and depends on just one global parameter of the 5D model.   
In \sref{4d}, I show that the same conclusions are reached in the framework of 4D effective theories describing the lightest fermionic KK modes.

Some consequences for higgs physics at the LHC are pointed out in \sref{d}.
The higgs production cross section can be significantly reduced, 
even down to $30\%$, for the range of parameters suggested by naturalness and electroweak precision tests.

%%%%%%%%%%%%%%%%%%%%%%%%%%%%%%%%%%%%%%%%%%%%%%%%%%%%%%%%%%%5
\section{Gluon fusion in 5D gauge-higgs unification}
\label{s.5d}
\setcounter{equation}{0} 
%%%%%%%%%%%%%%%%%%%%%%%%%%%%%%%%%%%%%%%%%%%%%%%%%%%%%%%%%%%%%%%%

This section contains a general discussion of the gluon-gluon-higgs amplitude in 5D models of gauge-higgs unification. 

The action for a 5D fermion multiplet reads
\beq
S_5 = \int d^4 x \int_{0}^{L} dy \sqrt{g} \left \{ 
 \ov \Psi (i \Gamma^N (D_N - i g_5 A_N)    - M) \Psi  \right \}
 \, . 
\eeq
Here $M$ is the bulk mass.   
$\Psi$ is a quark multiplet -  a triplet under color $SU(3)$.  
It is also charged under another group factor that contains the SM electroweak group so that it may contain several quark flavours with top, bottom or exotic quantum numbers. 
We will study this action in a general warped background with the line element $ds^2 = a^2(y) dx^2 - dy^2$. 

The fifth component of the gauge field may host physical degrees of freedom 
that,  in the KK picture, become massless (at tree level) 4D scalar fields. 
If the 5D gauge group is non-abelian, a vev of these scalar leads to spontaneous breaking of gauge symmetry.   
We single out one generator $T^\ha$  along which the vev  resides and 
we define the higgs boson field $h(x)$ as oscillations around the vev:   
\beq
A_5 \to  {a^{-2}(y) \over \left [\int_0^L a^{-2} \right ]^{1/2} } T^\ha (\ti v + h(x))  
\eeq
The normalization factors are chosen such as to make the higgs $h(x)$ canonically normalized in the KK picture. 

%The quadratic action in that background reads  
%\beq
%S_5 = \int d^4 x \int_{0}^{L} dy  \left \{ 
% i a^3 \ov \Psi \gamma^\mu \pa_\mu \Psi 
% -  a^4 \ov \Psi (\gamma^5 \hat D_y  - M  ) \Psi 
%  \right \}  \, .
%\eeq
%where $\hat D_y = \hat \pa_y - i g_5 \la A_5 \ra$ and $\hat \pa_y  = \pa_y + 2 a'/a$. 

We expand the fermions into the KK mass eigenstates 
\beq
\Psi_L(x,y) = f_{L,n}(\ti v,y) \Psi_{L,n}(x) 
\qquad
\Psi_R(x,y) = f_{R,n}(\ti v,y) \Psi_{R,n}(x) 
\eeq 
The profiles satisfy the equations of motion:
\bea
\left (\hat D_5 + M  \right  )f_{R,n} (\ti v,y)  &=& m_n a^{-1} f_{L,n} (\ti v,y)  
\nn
\left (-\hat D_5  + M  \right )f_{L,n}(\ti v,y)  &=& m_n a^{-1} f_{R,n} (\ti v,y) 
\eea 
and boundary conditions appropriate for the model. 
Furthermore, they satisfy the orthonormality conditions 
%\beq
%\int_0^L a^3(y)  f_{L,n}^\dagger(\ti v,y)  f_{L,m}(\ti v,y) = 
%\int_0^L a^3(y)  f_{R,n}^\dagger(\ti v,y)  f_{R,m}(\ti v,y) = \delta_{nm}
%\eeq 
from which follow the completeness relation 
\beq
\sum_n   f_{L,n}(\ti v,z) f_{L,n}^\dagger(\ti v,y) 
=
\sum_n    f_{R,n}(\ti v,z)  f_{R,n}^\dagger(\ti v,y) 
= a^{-3} (y) \delta (y - z) I,  
\eeq 
where $I$ is the identity matrix in the flavour space. 

The higgs boson couples to the fermionic eigenstates as 
$-y_{nm} h \ov \psi_{L,n} \psi_{R,m}$, where  the Yukawa couplings are given by
\beq
\label{e.hyc}
y_{nm} = - i g_5 {\int_0^L a^2 (y) f_{L,n}^\dagger(\ti v,y) T^\ha f_{R,n}(\ti v,y)   
\over  \left (\int_0^L a^{-2}(y)  \right )^{1/2} } . 
\eeq 
Although it is not obvious at this point, the couplings $y_{nm}$ are real. 

\begin{figure}
\label{f.ggh}
\centering{
\includegraphics{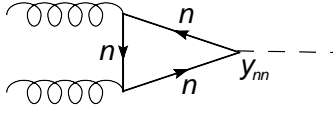}
}
\caption{
Contribution to the gluon-gluon-higgs amplitude of the $n$-th fermionic KK eigenstate.
Because the QCD coupling of quark eigenstates to gluons is diagonal and the eigenstates do not mix with each other, only diagonal Yukawa couplings are relevant.}
\end{figure}
The objective is to compute the one-loop contribution of the fermionic eigenstates to the gluon-gluon-higgs amplitude. 
The relevant Feynman diagram is depicted on \fref{ggh}.
In the SM, this amplitude is dominated by the top quark who has the largest Yukawa coupling to the higgs.
Assuming the higgs boson is light enough, $m_{\rm higgs} < 2 m_t$,   
the amplitude can be written as 
\beq
\cm_{\rm ggh}({\rm SM}) \approx f(\eps,p) {y_{tt} \over m_t} = f(\eps,p) {1 \over v}
\eeq  
where $f(\eps,p)$ depends on the momenta and polarizations of the incoming gluons and its precise form will be of no importance in the following. 
In the following we always assume that $m_{\rm higgs} < 2 m_n$, for the top quark as well as for all new physics quarks.
This is a safe assumption in the 5D pseudo-goldstone scenario: the higgs  potential is generated at one loop so that the higgs boson is expected to be light, not far from the present direct search limit.   
Thus, the amplitude can be approximated by  
\beq
\cm_{\rm ggh}({\rm 5D}) \approx f(\eps,p) \sum_n {y_{nn} \over m_n} 
\eeq 
where the sum goes over all heavy enough fermionic eigenstates (it includes the SM top quark, but not the bottom or any of the lighter quarks). 

Now we prove a remarkable sum rule.  
We first note that, using the equations of motion, the profiles can be  represented as:   
\bea
f_{R,n} (\ti v,y) &=& \Omega(y) a^{-2}(y) e^{-M y}
 \left (f_{R,n} (\ti v,0) + m_n  \int_0^y a(y') e^{M y'} \Omega^{-1}(y')f_{L,n} (\ti v,y') \right )
\nn
f_{L,n} (\ti v,y) &=& \Omega(y) a^{-2}(y) e^{M y}
 \left (f_{L,n} (\ti v,0) -  m_n \int_0^y a(y') e^{- M y'} \Omega^{-1}(y') f_{R,n} (\ti v,y') \right )
\nn
\eea 
where $\Omega(y)  = e^{ i g_5 \int_0^y \la A_5 \ra}$.  
Using the above expressions and the completeness relations one can derive  
\beq
\sum_n {y_{nn} \over m_n} = 
- i g_5    \left [\int_0^L a^{-2} \right ]^{1/2}  
\sum_n m_n^{-1} f_{L,n}^\dagger(\ti v,0) T^\ha f_{R,n}(\ti v,0)    . 
\eeq 
To clean up, we introduce the global symmetry breaking scale $f$, 
\beq
\label{e.f}
f = {\sqrt 2 \over g_5 (\int_0^L a^{-2})^{1/2}}. 
\eeq 
which in pseudo-goldstone higgs scenarios plays an analogous role as the pion decay constant in low-energy QCD.  
Furthermore, the sum in the last expression is related to the chirality flipping propagator in 4D momentum/5D position space.
More precisely, the propagator is defined by  
\beq
P_{RL}(y,z) = i \sum_n f_{R,n}(y) f_{L,n}^\dagger (z)    { m_n  \over p^2 - m_n^2} 
\eeq 
so that we rewrite 
\beq
\label{e.sum}
\sum_n {y_{nn} \over m_n} =  f^{-1}  {\tr}[\sqrt{2} T^\ha P_{RL}(0,0)]|_{p^2 \to 0}   
\eeq  
In the SM, the amplitude in the decoupling limit is proportional to $1/m_t$, which is the zero momentum limit of the top quark propagator. 
In 5D, this is generalized to the zero momentum limit of the UV boundary propagator.   
This kind of result could be expected from holography. 
In the end, the 5D set-up can be interpreted as a dual description of 4D fundamental quarks (living on the UV boundary) that mix with fermionic composite operators from a strongly coupled sector \cite{copo}.  
It should be underlined, however, that \eref{sum} is not a "holographic prescription", but a rigorous results derived from the 5D formalism.  

Finally, we define $R_g^{1/2}$ as the ratio of the gluon-gluon-higgs amplitude in the 5D model to that in the  SM (with $m_h < 2 m_t$). 
The ratio is given by the following simple expression 
\beq
\label{e.rg} % R general
R_g^{1/2} = {v \over f} \left ( 
{\tr}[\sqrt{2} T^\ha P_{RL}(0,0)]|_{p^2 \to 0} - \sum_{\rm light} {y_{nn} \over m_n}
\right ) 
\eeq

The methods of computing fermionic propagators in a warped background are reviewed in \aref{fp}.
In the next section we apply \eref{rg} in the context of realistic 5D models of pseudo-goldstone higgs.

%%%%%%%%%%%%%%%%%%%%%%%%%%%%%%%%%%%%%%%%%%%%%%%%%%%%%%%%%%%%%%%%%%%%%%%%%%%  
\section{SO(5) models}
\label{s.so5} 
\setcounter{equation}{0}
%%%%%%%%%%%%%%%%%%%%%%%%%%%%%%%%%%%%%%%%%%%%%%%%%%%%%%%%%%%%

We apply the general methods outlined in the previous section in the context of 5D models with the electroweak group embedded  in $SO(5) \times U(1)_X$ \cite{agcopo}.  
This is the simplest set-up that accommodates the correct Weinberg angle and the custodial symmetry.  
The latter is indispensable in 5D warped models in order to keep the Peskin-Takeuchi  T parameter under control \cite{agdema,caposa}. 
$SO(5)$ has 10 generators: three form the $SU(2)_L$ subgroup (identified with the standard model $SU(2)$), 
another three form the $SU(2)_R$ subgroup (identified with the custodial symmetry) and the remaining four generators $T_C^b$ belong to the $SO(5)/SO(4)$ coset.  
Some useful facts  about $SO(5)$ are collected  in \aref{g}. 

The $SO(5)$ gauge symmetry is reduced on the boundaries by imposing Dirichlet boundary conditions for some of the generators. 
The surviving gauge symmetry on the UV brane is $SU(2)_L \times U(1)_Y$, the hypercharge being a linear combination of the $T_R^3$ and $U(1)_X$ generators.   
On the IR brane, the symmetry is reduced down to $SO(4) = SU(2)_L \times SU(2)_R$.
The four generators from the  $SO(5)/SO(4)$ coset have Dirichlet boundary conditions on both branes, so that the fifth components of the corresponding gauge fields hosts scalar fields identified with  the  SM higgs doublet. 
The vev is chosen  along the $T_C^4$ generator. 
The electroweak breaking scale is $v = f \sin (\ti v/f)$, where $f$ is defined in \eref{f}.

There are several options for embedding the third generation quarks into $SO(5)$ multiplets. 
The first model we consider here is a variation on that introduced in ref. \cite{agcopo}.  
The top and bottom quarks are embedded into two 5D quarks in the {\it spinorial} representation $\bf 4_{1/6}$: 
\beq
Q_1 =  ( q_1, \ q_1^c)   = ( t_1, \ b_1, \ t_1^c, \  b_1^c )   
\qquad 
Q_2 =  (q_2, \  q_2^c) = (  t_2, \ b_2, \  t_2^c, \  b_2^c ) 
\eeq  
The IR boundary conditions are the same for $Q_1$ and $Q_2$,  
\beq
q_{R,i}(L) =  q_{L,i}^c(L) =  0 
\eeq
The  UV boundary conditions are chosen as\footnote{The peculiar UV boundary conditions for the electroweak doublets $q_i$ can be realized  by mixing the linear combination $\bar \theta_1 q_{1,L} + \bar \theta_2 q_{2,L}$  with a UV boundary fermion $\ti q_R$ through a boundary mass term, and taking the boundary mass to infinity.}
\bea 
\theta_2 q_{1,R}(0) - \theta_1 q_{2,R}(0)  &=& 0 
\nn 
\bar \theta_1 q_{1,L}(0) + \bar \theta_2 q_{2,L}(0)  &=& 0 
\nn 
t_{1,L}^c (0) = t_{2,R}^c (0)    &=& 0 
\nn
b_{1,R}^c (0) = b_{2,L}^c (0)    &=& 0 
\eea 
The KK towers include two quark eigenstates that become massless in the limit of no electroweak breaking. 
These are identified with the SM top and bottom quarks. 
The mass splitting between the top and the bottom quark can achieved 
if $|\theta_1/\theta_2| \ll 1$ or if $M_1 > M_2$, 
in which case 
$m_b^2/m_t^2 \approx |\theta_1/\theta_2|^2 \int_0^L a^{-1} e^{-2 M_1 y}/\int_0^L a^{-1} e^{-2 M_2 y}$. 
 
In this scenario, 
all the quark eigenstates couple to the higgs boson and contribute to the gluon-gluon-higgs amplitude.   
We first compute the contribution of the top quark tower. 
Using \eref{sum}, 
\beq
\sum_{\rm top} {y_{nn} \over m_n}  
=  {1 \over 2 f} \left ( 
P_{RL}^{t_1 t_1^c } (0,0)   
+
P_{RL}^{t_1^c t_1 }(0,0) 
+
P_{RL}^{t_2 t_2^c }(0,0)   \right)|_{p^2 = 0}
\eeq  
We compute the UV propagators using the algorithm outlined in  \aref{fp}. 
The final result is very simple, 
\beq
\sum_{\rm top} {y_{nn} \over m_n} =  {\cos(\ti v/f) \over f \sin(\ti v/f)} 
\eeq  
The same result is obtained for $\sum_{\rm bottom} {y_{nn} \over m_n}$.
In that case, however, the sum is almost entirely dominated by the lightest bottom quark contribution. 
More precisely, starting from \eref{hyc} one finds $y_{bb}/m_b =f^{-1} \cot(\ti v/f) + \co(m_b^2)$.
Thus, by \eref{rg}, the bottom quark tower does not contribute significantly to the gluon-gluon-higgs amplitude.
We conclude that in this model based on the spinorial representation  
\beq
\label{e.rs} % R spinorial 
R_g^{1/2} \approx \cos(\ti v/f) + \co(m_b^2/\mkk^2) 
\approx \sqrt{1 - v^2/f^2}.
\eeq

\vspace{.5cm}

In the second model we consider, the third generation is embedded in the {\it fundamental} $SO(5)$ representation \cite{codapo}.
We consider two 5D quarks $Q_1$, $Q_2$ transforming as  $\bf 5_{2/3}$, $\bf 5_{-1/3}$, respectively. 
The quantum numbers are embedded into the fiveplet as  
\beq
Q_1 = {1 \over \sqrt{2}} \bvec 
 \chi +  b_1 \\ i (\chi - b_1), 
 \\ t_1 + \ti t \\ i (t_1- \ti t)  \\  \sqrt{2} t^c 
 \evec   
\qquad 
Q_2 =   {1 \over \sqrt{2}} \bvec 
 \rho +  t_2 \\ i (\rho - t_2), 
 \\ b_2 + \ti b \\ i (b_2- \ti b)  \\   \sqrt{2} b^c 
 \evec   
\eeq
The exotic quarks $\chi$ and $\rho$ have electric charges $5/3$ and $-4/3$, respectively.  
The upper four component in each multiplet can be collected into bi-doublets of $SU(2)_L \times SU(2)_R$: 
$\Phi_i = (q_i,\ti q_i)$, where $q_i = (t_i,b_i)$, 
$\ti q_1 = (\chi, \ti t)$, $\ti q_2 = (\ti b,\rho)$. 
The fact that the top-bottom $SU(2)_L$ doublet is embedded into a bi-doublet 
protects the SM Zbb vertex against dangerous corrections \cite{agcoda}, 
which provides rationale for this more complicated model. 

The IR boundary conditions are chosen as  
\beq
\Phi_{R,1}(L) =  \Phi_{R,2}(L) =  0 
\qquad
t_L^c(L) = b_L^c (L)  = 0 . 
\eeq
The UV boundary conditions are  
\bea
\theta_2 q_{1,R}(0) - \theta_1 q_{2,R}(0)  &=& 0 
\nn 
\bar \theta_1 q_{1,L}(0) + \bar \theta_2 q_{2,L}(0)  &=& 0 
\nn 
t_{L}^c(0) = b_{L}^c(0)    &=& 0 
\nn
\ti q_{1,L}(0) =  \ti q_{2,L}(0)    &=& 0 . 
\eea 

The exotic quarks do not couple to the higgs boson at all. 
Furthermore, using the same arguments as previously, one can prove that the bottom quark tower contribution to gluon-gluon-higgs amplitude is suppressed by the small bottom quark mass.
As for the top quark tower contribution, we compute 
\beq
\sum_{top} {y_{nn} \over m_n} =
 {1 \over \sqrt 2 f} \left ( 
 - P_{RL}^{t^c t_1} + P_{RL}^{t^c \ti t}
\right )|_{p^2 = 0} =  {\cos(\ti v/f) \over f \sin(\ti v/f)} 
\eeq   
This is the same result as that in obtained in the spinorial model. 
However, the present set-up is incomplete as it stands, because it does not have a correct electroweak breaking vacuum.
In the language of ref. \cite{aa}, 
the spectral function corresponding to the top quark tower is 
$\rho \approx S_{M_1}(L) S_{-M_1}(L) -  (1/2) \sin^2(\ti v/f)$ and the minimum of the higgs potential falls at  
$\sin^2(\ti v/f)  = 1$, which implies that the model is equivalent, in practice, to a higgsless theory \cite{gigrpo,bary}. 
In order to achieve $\sin^2(\ti v/f)  \ll 1$, as suggested by electroweak precision tests, 
we need to introduce the so-called shadow multiplet, whose role is to produce a  quartic term, $\sin^4(\ti v/f)$, in the spectral function \cite{agcopo,mewa,aa_susy}. 
A shadow multiplet, is another 5D quark  that has no light modes in the limit of no electroweak breaking; 
massless modes appear however for maximal electroweak breaking. 
Here we consider a 5D quark $S$ transforming as $\bf 5_{2/3}$. 
It contains a bi-doublet $\Phi^s$ and a singlet $t^s$ with the boundary conditions chosen as
\beq
\Phi_R^s(0) = t_L^s(0) =  \Phi_L^s(L) = t_R^s(L) = 0 .
\eeq 
In this set-up, the contribution of the new shadow top quark tower to the gluon-gluon-higgs amplitude is given by 
\beq  
\sum_{\rm shadow} {y_{nn} \over m_n} = - {\sin (\ti v/f)  \over f \cos (\ti v/f)}
\eeq
The minus sign implies that the shadow tower interferes destructively with that of the top. 
At the end of the day we find  that in this model based on the fundamental representation   
\beq
\label{e.rf}
R_g^{1/2}  \approx {\cos (2 \ti v/f)  \over \cos (\ti v/f)}
\approx {1 - 2 v^2/f^2 \over \sqrt{1 - v^2/f^2}}
. 
\eeq 

The shadow multiplet can mix with $Q_1$ and $Q_2$  via IR boundary mass terms. 
Including these mass terms  does not change the above result. 
If another shadow multiplet is included, for example $\bf 5_{-1/3}$ as in \cite{codapo}, it would further diminish the amplitude.

One could repeat the same procedure for other gauge-higgs unification models, e.g. for $SO(5)$ models with the the third generation embedded in the adjoint representation or for models based on the $SU(3)$ gauge group. 
The recurring feature of the gluon-fusion amplitude is that the ratio $R_g$ depends only on $v/f$ and is not sensitive to the details of the KK quark spectrum.  
Furthermore, in all cases one finds $R_g^{1/2} < 1$.

%%%%%%%%%%%%%%%%%%%%%%%%%%%%%%%%%%%%%%%%%%%%%%%%%%%%%%%%%%%%%%%%%%%%%%%%%%%  
\section{4D effective description}
\label{s.4d} 
\setcounter{equation}{0}
%%%%%%%%%%%%%%%%%%%%%%%%%%%%%%%%%%%%%%%%%%%%%%%%%%%%%%%%%%%%

The results for the gluon-gluon-higgs amplitude that were obtained in 5D, can be reproduced in a 4D framework. 
In the spirit of refs. \cite{ami,rob,bary}, one can construct a 4D model that mimics the low-energy dynamics of 5D gauge-higgs models. 
The gauge group is the SM one, $SU(3)_C \times SU(2)_L \times U(1)_Y$, while the bulk gauge group of the 5D set-up is realized as an approximate {\em global} symmetry, which is spontaneously broken by a vev of a scalar field. 
The pseudo-goldstone higgs resides in that scalar field. 
The quark sector includes the standard model quarks and a finite number of vector-like quarks. The 4D set-up can be considered as an effective low-energy description of the 5D gauge-higgs unification setup (or some 4D strong dynamics), with the cut-off identified  with the resonance scale. 
%We will show that in such setting the 5D results for the glue-glue-higgs amplitude can be reproduced. 
%, provided we implement the cancellation of quadratic divergences.   

In 4D,  Yukawa couplings of the higgs boson to the quarks are  given by $y_{nm} = {\pa M_{nm} \over \pa \ti v}$, where $M$ is the quark mass matrix and $\ti v$ is the higgs vev.   
Since we deal with  a finite number of quarks, 
it is more handy to adopt another sum rule to compute the quark contributions to the gluon-gluon-higgs amplitude.  We can rewrite: 
\beq
\sum {y_{nn} \over m_n} = \tr(y M^{-1}) =
\tr  \left ( {\pa M \over \pa \ti v} M^{-1} \right )
=  {\pa \tr \log M \over \pa \ti v}  
=  {\pa \log \det M \over \pa \ti v} 
\eeq   
Thus, the gluon-gluon-higgs amplitude can be related to the determinant of the quark mass matrix \cite{elgano},   
\beq
\label{e.rsmart}
R_g^{1/2} =  v  {\pa \over \pa \ti v} \log \det M(\ti v) 
\eeq 
In fact, this formula can be also applied to 5D gauge-higgs models, in spite of the fact that $\det M(\ti v)$ diverges in 5D due to infinite multiplicity of KK states.\footnote{This point was clarified in private discussions with Csabi Csaki and Andy Weiler.}
In the gauge-higgs case, we should replace $\det M(\ti v)$ in \eref{rsmart} with $\rho^{1/2}(0)$, 
where $\rho(p^2) \equiv \det (-p^2 + m_n^2)$ is the spectral function that can be computed by solving the 5D equations of motion \cite{aa,aa_susy}.  
This method yields the same results as the position propagator method used in \sref{5d}.  

The 4D effective description of the 5D model based on the $SO(5)$ spinorial representation is the following \cite{bary} (see also \cite{ami}). 
Left-chirality quarks include two $SU(2)_L$ doublets $q_{L,i} = (t_{L,1},b_{L,i})$, one singlet top $T_L^c$ and one singlet bottom $B_{L}^c$.  Right-chirality quarks include two top singlets $t_{R,i}^c$, two bottom singlets $b_{R,i}^c$ and one doublet  $Q_{R} = (T_R,B_R)$. 
This amounts to three top Dirac states, another three bottom Dirac states and no exotics. 
The quarks are collected into $SO(5)$ spinors as follows 
\beq
\Psi_L = \bvec
 q_{L,1} 
 \\
-i  B_L^c 
 \\
-i  T_L^c \evec 
 \qquad
\Psi_R = \bvec
 Q_{R} 
 \\
i  b_{R,1}^c 
 \\
i t_{R,1}^c \evec  
\eeq 
The global $SO(5)$ symmetry is spontaneously broken by a scalar field transforming as $\bf 5_0$. 
The pseudo-goldstone higgs is embedded in that scalar as $\Phi_M \to f(0,0,0,\sin (\ti v/f),\cos(\ti v/f))$. 
The Yukawa couplings  respect  $SO(5)$.
The global symmetry is  explicitly yet softly broken by vector-like mass terms,  
\beq
- \cl_{yuk} =  y \ov\Psi_L \Gamma^M \Phi_M \Psi_{R} 
+ f \lambda_q   \ov q_{L,2} Q_{R}   
+ f \lambda_t   \ov T_L^c t_{R,2}^c
 +f \lambda_b   \ov B_L^c b_{R,2}^c 
+ \hc 
\eeq
The resulting mass terms for the top quark can be represented in the matrix, 
\beq
- \cl_{mass} =  (\ov t_{L,1},\ov t_{L,2},\ov {T_L^c}) M \bvec t_{R,1}^c \\ t_{R,2}^c \\  T_R \evec
\eeq
\beq
M = f \left ( \ba{ccc}
y  \sin (\ti v/f)  &  0   & y \cos (\ti v/f) 
\\
0  & 0 & \lambda_q
\\
y \cos (\ti v/f)  & \lambda_t &  - y \sin (\ti v/f)
 \ea \right )   
\eeq 
The trace of the mass matrix squared does not depend on $\ti v$, which implies that the top quarks generate no  quadratically divergent corrections to the higgs mass parameters at one loop. In fact, this model is supersoft; the logarithmic divergences cancel as well.  The determinant of the mass matrix is proportional to $\lambda_q \lambda_t y \sin (\ti v/f)$. 
From \eref{rsmart} we find  
\beq
R_g^{1/2}  = \cos (\ti v/f) 
\eeq
in accord with \eref{rs}.

\vspace{0.5cm}

Another model provides an effective description to the 5D model based on the $SO(5)$ fundamental representation \cite{bary}. 
Left-chirality quarks include two $SU(2)_L$ doublets $q_{L,i} = (t_{L,i},b_{L,i})$, one singlet $T_L^c$ and one exotic (hypercharge $7/6$) doublet $\ti Q_{L} = (\chi_L, \ti T_L)$.  
Right-chirality quarks include two singlets $t_{R,i}^c$, one doublet $Q_R = (T_R,B_R)$ and one exotic doublet $\ti Q_{R} = (\chi_R,\ti T_R)$. 
This makes four Dirac quarks with top quantum numbers, one massless chiral bottom quark, one Dirac bottom quark and one exotic quark with charge $5/3$. 
More structure is needed to give mass to the bottom quark, but we do not elaborate on  it here. 
The quark fields are collected into $\bf 5$'s as follows  
\beq
\Psi_L = \bvec
 {1 \over \sqrt{2}} (b_L + \chi_L) 
 \\
 - {i \over \sqrt{2}} (b_L - \chi_L) 
 \\
 {i \over \sqrt{2}} (t_{L,1} - \ti T_L) 
 \\ 
 {1 \over \sqrt{2}} (t_{L,1} + \ti T_L) 
 \\
 T_L^c \evec 
\qquad 
\Psi_R = \bvec
 {1 \over \sqrt{2}} (B_R + \chi_R) 
 \\
 - {i \over \sqrt{2}} (B_R - \chi_R) 
 \\
 {i \over \sqrt{2}} (T_R - \ti T_R) 
 \\ 
 {1 \over \sqrt{2}} (T_R + \ti T_R) 
 \\
 t_{R,2}^c \evec  
\eeq 
   
We write down  $SO(5)$ symmetric Yukawa couplings and soft-breaking  vector-like mass terms,   
\bea 
- \cl_{yuk} =  & y_1 \ov\Psi_L \Phi t_{R,1}^c + y_2 \ov{T_L^c} \Phi^\dagger \Psi_R  
\nl 
f \lambda_1   \ov q_{L,1} Q_{R}  
+ f \lambda_2   \ov q_{L,2} Q_{R}  
%+ f \lambda_3   \ov T_L^c t_{R,2}^c   
+ f \lambda_\chi   \ov {\ti Q_L} \ti Q_{R} + \hc 
\eea
Some mass terms allowed by gauge symmetries,  e.g.  $T_L^c t_{R,2}^c$,  are omitted because they would violate the softness of the global symmetry breaking.
The mass matrix for the top quarks: 
\beq
- \cl_{mass} = (\ov t_{L,1},\ov t_{L,2}, T_L^c, \ov {\ti T_L}) M \bvec t_{R,1}^c \\ t_{R,2}^c \\ T_R \\ \ti T_R \evec
\eeq
\beq
M = f \left ( \ba{cccc}
y_1 \sin (\ti v/f)  & 0  & \lambda_1 & 0 
\\
0  & 0  & \lambda_2 & 0
\\ 
\sqrt{2} y_1 \cos (\ti v/f)  & \sqrt{2} y_2 \cos (\ti v/f) & y_2 \sin (\ti v/f) & y_2 \sin (\ti v/f)
\\
y_1 \sin (\ti v/f)  & 0 & 0 & \lambda_\chi
 \ea \right )   
\eeq 
The determinant is now proportional to  $\sin (2 \ti v/f)$, rather than to $\sin (\ti v/f)$ as in the spinorial realization. 
In consequence,     
\beq
R_g^{1/2}  = {\cos (2 \ti v/f)  \over \cos (\ti v/f)}
\eeq 
as in \eref{rf}.

%%%%%%%%%%%%%%%%%%%%%%%%%%%%%%%%%%%%%%%%%%%%%%%%%%%%%%%%%%%%%%%%%%%%%%%%%%%  
\section{Discussion}
\label{s.d} 
\setcounter{equation}{0}
%%%%%%%%%%%%%%%%%%%%%%%%%%%%%%%%%%%%%%%%%%%%%%%%%%%%%%%%%%%%%%%%%%%%%%%%%%%%%%%%
 
I presented an analysis of the gluon-gluon-higgs amplitude in two models with a pseudo-goldstone  higgs boson realized in 5D $SO(5)$ gauge-higgs unification.  
In both cases the higgs production cross-section is {\it suppressed} with respect to the SM result. 
%One can argue \cite{ian} that the suppression is a generic feature of models that solve the hierarchy problem.
Suppression effects were also concluded in a closely related framework of  little higgs \cite{bob} (in gauge-higgs, the suppression was alluded to in ref. \cite{caposa}). 
I did not  prove that pseudo-goldstone higgs scenarios with an enhanced higgs production cross section  do not exist. 
In fact, one can write down a somewhat stretched counter-example where $R_g^{1/2} < -1$ due to contributions of several shadow multiplets, so that the cross section would be enhanced.       
The fact is, however, that the simplest setups consistent with naturalness and electroweak precision tests always predict suppression of the higgs production rate.
In contrast, enhancement can be achieved in the parameter space of 5D models in which the higgs potential is not protected, for example in UED  \cite{pet}, or in the warped models based  on the $SU(2)_L \times SU(2)_R$ group \cite{djmo}.    
Enhancement could also be achieved in the MSSM although, in that case, in the most interesting parameter space region with the minimal electroweak fine-tuning one also finds a suppression \cite{delo}. 

%In the pseudo-goldtone scenario  there are good reasons to expect that  the top and new physics quarks are much heavier than the higgs. 
The most interesting result obtained in this paper is that, if $m_h < 2 m_n$ for the top quark and all new physics quarks, the suppression factor $R_g$ in 5D models depends very little on the details of the spectrum.
%, which is fixed dynamically by minimization of the higgs potential.  
One could expect that the result depends on the individual masses of the vector-like quarks, since  
the Yukawa coupling of the top quark can be substantially modified in the presence of fairly light new physics states. 
However, summing up the contributions of the whole KK tower  leaves only the dependence on $v/f$ -- the ratio of the  electroweak breaking scale to the global symmetry breaking scale. 
Thus, the prediction for the gluon-fusion amplitude is a very robust feature of the gauge-higgs models.

The suppression factor $R_g$ depends also on the embedding of the third generation into $SO(5)$ multiplets,  of which several options exist in the literature \cite{agcopo,codapo}.   
Given that we identify the embedding by observing some of the fermionic resonances, a precise enough measurement of the higgs production cross section could provide a simple way to determine the scale $f$.
The latter is very important phenomenologically, as it controls the growth of the longitudinal gauge boson scattering amplitude below the resonance scale \cite{gigrpo};
knowing the scale $f$ would provide an answer to the question if strong WW scattering occurs at the energies accessible at the LHC.

There are two theoretical arguments concerning the actual value of $f$, 
that hint towards a  different range.
One one hand, the little hierarchy problem suggests $f$ should not be too large because  
the fine-tuning needed to achieve $v/f \ll 1$ is proportional to $v^2/f^2$  \cite{agcopo}. 
On the other hand, electroweak precision tests suggest a larger value, 
as $v/f$ of order unity corresponds to an effectively heavy higgs \cite{bary}, which is disfavoured by electroweak data. 
Furthermore, we expect $f \geq 500 \gev$, since a smaller value implies the existence of vector resonances with masses below $3 \tev$ (to unitarize the WW scattering), which is disfavoured by electroweak precision data.  
With the above facts in mind, I pick up two benchmark points 
$f = 500 \gev$, $f = 1 \tev$ corresponding to the fine-tuning of order $25\%$ and $5\%$,  respectively,
and to the effective higgs mass $250 \gev$ and $145 \gev$ (for the true higgs mass $115 \gev$).   
The suppression factor in the two models we studied is given by   
\bc
\btb[h]{|c|c|c|}
\hline
  & ${\bf 4}$ & ${\bf 5}$
\\
\hline
$f = 500 \gev$ & $R_g =75\%$ & $R_g = 35 \%$ 
%\hline
\\
$f = 1000 \gev$ & $R_g =95\%$ & $R_g = 82 \%$ 
\\
\hline
\etb
\ec
For a reasonable choice of parameters, the suppression can be particularly large in the model based on the fundamental ($\bf 5$) representation, the one that is favoured by the measurements of the $Z \bar b b$ vertex.   
Using the diphoton channel at the LHC, the theoretical estimate of the higgs production cross section may be confronted with experiment with $\sim 10\%$ accuracy \cite{babis}. 
Thus, the suppression effect due to the pseudo-goldstone nature of the higgs boson should be confirmed at the LHC in most of the interesting parameter space.  

Note that the photon-photon-higgs amplitude is suppressed too; that amplitude is dominated by a loop of W boson whose coupling to the higgs boson is suppressed by $\cos(\ti v/f)$ \cite{gigrpo}. 
However, other decay amplitudes, for example $h \to \bar b b$ are typically suppressed too.    
The modification of the branching ratios depends on the embedding of the SM fields into multiplets of the bulk gauge group.
In the two models studies in \sref{so5} the branching ratios are not changed.

If   gauge-higgs unification is realized in nature, higgs boson searches at the LHC may be more challenging than previously assumed. 
The suppression of the higgs production cross-section predicted by theses models may obstruct a quick discovery of the higgs boson. 
On the positive side, if fairly light vector-like quarks are present, decays of these quarks  may enhance the higgs production \cite{caposa}.
Finally, not discovering the higgs boson after $30 \ {\rm fb}^{-1}$ of data should not be considered discouraging but rather a hint that the pseudo-goldstone mechanism is at work ;-).

\section*{Acknowledgements}

I am grateful to Babis Anastasiou for inspiring this work and to Roberto Contino for clarifications on $SO(5)$ models. Thanks to Babis Anastasiou, Ian Low and Jose Santiago for comments on the manuscript and  for pointing out blunders in an earlier version.
I also profited from post-v1 discussions with Csaba Csaki, Ami Katz, Maria Krawczyk, Eduardo Pont\'on, Carlos Wagner, Andy Weiler and Neal Weiner. 

I am partially supported by the European Community Contract MRTN-CT-2004-503369 for the years 2004--2008 and by  the MEiN grant 1 P03B 099 29 for the years 2005--2007. 

\renewcommand{\thesection}{Appendix \Alph{section}}
%\appendix 
\renewcommand{\theequation}{\Alph{section}.\arabic{equation}} 
\setcounter{section}{0} 
\setcounter{equation}{0} 

%%%%%%%%%%%%%%%%%%%%%%%%%%%%%%%%%%%%%%%%%%%%%%%%%%%%%%%%%%%%%%%%%%%%%%%%%%%%% 
\section{Fermionic propagator in gauge-higgs background} 
\label{a.fp} 
%%%%%%%%%%%%%%%%%%%%%%%%%%%%%%%%%%%%%%%%%%%%%%%%%%%%%%%%%%%%%%%%%%%%%%%%%%%%%%

The propagator can be equivalently defined either as an inverse of the kinetic operator in the lagrangian 
or, as in the following, as a sum of KK propagators weighted by profiles. 
We define the mixed (4D momentum/5D position space) fermionic propagators by 
\bea 
P_{LL}(y,z) = i\sum_n { f_{L,n}(y)  f_{L,n}^\dagger (z)  \over p^2 - m_n^2} 
&\qquad&
P_{RR}(y,z) = i\sum_n { f_{R,n}(y)  f_{R,n}^\dagger (z) \over p^2 - m_n^2} 
\nn
P_{LR}(y,z) = i\sum_n { m_n f_{L,n}(y)  f_{R,n}^\dagger (z) \over p^2 - m_n^2} 
&\qquad&
P_{RL}(y,z) = i\sum_n { m_n f_{R,n}(y)  f_{L,n}^\dagger (z) \over p^2 - m_n^2} 
\eea 
The propagators are matrices in the flavour space and satisfy the coupled differential equations 
\bea
\label{e.pe} % propagator equations 
i a^{-3}\delta (y - z) &=& 
 p^2 P_{LL} - a (\hat D_y + M) P_{RL} 
\nn
i a^{-3}\delta (y - z) &=& 
p^2 P_{RR} - a (-\hat D_y + M) P_{LR} 
\nn
0 &=& 
P_{LR} - a (\hat D_y + M) P_{RR} 
\nn
0 &=& 
 P_{RL} - a (-\hat D_y + M) P_{LL} ,   
\eea 
which can be derived using the equations of motions and the completeness relations for the profiles.  
To solve these equations we introduce auxiliary (hatted) propagators, 
separately for $y < z$ and for $y > z$   
\beq
P_{LL}^<(y,z)
 = a^{-2}(y) e^{M y} \Omega(y) \hat P_{LL}^<(y,z)
\qquad
P_{RR}^<(y,z) = a^{-2}(y) e^{-M y} \Omega(y) \hat P_{RR}^<(y,z)
\eeq 
\beq
P_{LL}^>(y,z)
 = a^{-2}(y) e^{M y} \bar \Omega(y) \hat P_{LL}^<(y,z)
\qquad
P_{RR}^>(y,z) = a^{-2}(y) e^{-M y} \bar \Omega(y) \hat P_{RR}^>(y,z)
\eeq 
where the Wilson rotation matrices act in the flavour space and are given by 
\beq
\Omega(y) = e^{i g_5 \int_0^y \la A_5 \ra} 
\qquad
\bar \Omega(y) = e^{i g_5 \int_L^y \la A_5 \ra} .  
\eeq 
The hatted propagators satisfy the second order differential equations:   
\beq
\label{e.peoml} % propagator equations of motion  left 
\left [ a e^{-2 M y} \pa_y ( a e^{2 M y} \pa_y ) + p^2 \right ] \hat P_{LL} = 0
\eeq
\beq 
\label{e.peomr} % propagator equations of motion right 
\left [ a e^{2 M y} \pa_y ( a e^{-2 M y} \pa_y ) + p^2 \right ] \hat P_{RR} = 0
\eeq
that are valid for both $P^<$ and $P^>$ as long as $y \neq z$. 
The matching conditions at $y = z$ follow from \eref{pe}, 
\bea 
\label{e.pg} % glueing 
\Omega(L) \hat P_{LL}^<(z,z) &=&  \hat P_{LL}^>(z,z)
\nn 
 \Omega(L) \pa_y \hat P_{LL}^<(y,z)|_{y=z} &=&   \pa_y \hat P_{LL}^>(y,z)|_{y=z} 
- i a^{-3}(z) e^{- M z} \bar\Omega^{-1}(z)
\nn
\Omega(L) \hat P_{RR}^<(z,z) &=&  \hat P_{RR}^>(z,z)
 \nn   
\Omega(L) \pa_y \hat P_{RR}^<(y,z)|_{y=z} &=&  \pa_y \hat P_{RR}^>(y,z)|_{y=z} 
- i a^{-3}(z) e^{M z} \bar \Omega^{-1}(z) 
\eea
The equations of motion \erefn{peoml} and \erefn{peomr} together with the matching conditions \erefn{pg} fully determine the chirality-diagonal propagators, once the boundary conditions are specified.   
The chirality-flipping propagators can be calculated from the diagonal ones.   
\beq
\label{e.cfp} % chirality flip propagators
P_{LR}^<(y,z) =  a^{-1} (y) e^{-M y} \Omega(y)\pa_y \hat P_{RR}^<(y,z)
\quad
P_{RL}^<(y,z) = - a^{-1} (y) e^{M y} \Omega(y) \pa_y \hat P_{LL}^< (y,z)
\eeq 
\beq
P_{LR}^>(y,z) =  a^{-1} (y)e^{-M y} \bar \Omega(y)\pa_y \hat P_{RR}^>(y,z)
\quad 
P_{RL}^>(y,z) = - a^{-1} (y)e^{M y} \bar \Omega(y) \pa_y \hat P_{LL}^> (y,z)
\eeq 

\vspace{0.5cm}

As an example, we compute the propagators in a simple toy-model.
Consider a 5D quark field $Q$ with a bulk mass $M$, transforming in the spinorial  $\bf 4$ representation of $SO(5)$ and having the $U(1)_X$ charge equal to $1/6$.
The fermion contains fields with quantum numbers of the SM top and bottom quarks embedded as follows  
\beq
Q =  ( q, \ q^c)   = ( t, \ b, \ t^c, \  b^c )   
\eeq 
The boundary conditions are  
\beq
\label{e.tmbc} % toy model boundary conditions 
q_R(0) = q_L^c(0)  = 0 \qquad q_R(L) = q_L^c(L)  = 0 . 
\eeq
This model is not realistic for several reasons, one being that it predicts degenerate top and bottom quarks, but it is simple enough to serve the illustration purpose.

First, following ref. \cite{aa}, we denote two independent solutions of \eref{peoml}  as $C_M(y)$ and $S_M(y)$. 
We pick up these solutions such that they satisfy  $C_M(0) = 1$, $C'_M(0) = 0$, $S_M(0) = 0$, $S'_M(0) = p$. 
The notation is to stress the similarity to  the familiar sines and cosines (to which these functions reduce for a flat warp factor and $M=0$). 
The warped generalization of $\sin' = \cos$ is 
$S_{M}'(y)   =   p a^{-1}(y) e^{-2 M y}  C_{-M}(y)$, 
$C_{M}'(y)   = - p a^{-1}(y) e^{-2 M y}  S_{-M}(y)$. 
The generalization of $\sin^2 + cos^2 = 1$ is the Wronskian $S_M(y) S_{-M}(y)+C_M(y) C_{-M}(y) = 1$.  
The explicit form of these solutions is of no relevance here; 
the only important property is that, at small momenta,  we can approximate $C_M = 1 - \co(p^2)$, 
$S_M = p \int_0^y a^{-1}(y') e^{- 2 M y'} + \co(p^3)$.
We also introduce the combinations 
$\bar C_M(y) = a(L) e^{2 M L} p^{-1} [ C_M(y) S'_M(L)  - S_M(y) C'_M(L)]$, 
$\bar S_M(y) = a(L) e^{2 M L} [ -C_M(y) S_M(L)  + S_M(y) C_M(L) ]$,  
that satisfy simple boundary conditions on the IR brane: 
$\bar C_M(L) = 1$, $\bar C'_M(L) = 0$, $\bar S_M(L) = 0$, $\bar S'_M(L) = p$. 

Armed with this formalism, we write the hatted propagators as 
\bea
\hat P_{LL}^{<qa} = C_{M} (y) c_L^{<qa}(z)  
&\quad&
\hat P_{LL}^{>qa} = \bar C_{M} (y) c_L^{>qa}(z)  
\nn
\hat P_{LL}^{<q^c a} = S_{M} (y) c_L^{<q^c a}(z)  
&\quad&
\hat P_{LL}^{>q^c a} = \bar S_{M} (y) c_L^{>q^c a}(z)
\eea  
where $a = q,q^c$. 
This form is dictated by the boundary conditions \erefn{tmbc}. 
Now we plug this into the matching equation \erefn{pg} and solve for the coefficient $c^{ab}(z)$. 
In particular, we find    
\beq  
c_L^{q^c q}(0) = {\sin (\ti v/f) \over 2 p} 
{1 - 2 S_{M}(L) S_{-M}(L) 
\over 
S_{M}(L) S_{-M}(L) - \sin^2 (\ti v/2f)
}  
\eeq 
For our purpose, we need only this coefficients since 
$\tr  T^\ha P_{RL}(0,0) = P_{RL}^{q^c q}(0,0)/\sqrt{2}$ and, from \eref{cfp}, 
$P_{RL}^{q^c q}(0,0) = - p  c_L^{q^c q}(0)$.
Poles of the propagator occur at  $S_{M}(L) S_{-M}(L) - \sin^2 (\ti v/2f)$, which determines
the fermionic resonance spectrum. 
The limit $p^2 \to 0$ is achieved by setting $S_{\pm M} \to 0$. 
We find 
\beq
\tr [\sqrt{2} T^\ha P_{RL}(0,0)] =  
2 {\cos(\ti v/2f) \over  2 \sin(\ti v/2f)}  
\eeq 
We have exposed the factor of two to stress that the formula includes contributions of degenerate top and bottom KK towers. 
Thus, in this toy-model, 
the contribution of top quark tower is  equal to $R^{1/2} = \cos^2(\ti v/2f)$ that of the SM top quark. 

In the realistic models, the propagator is determined according to the same algorithm; the computation is just a tad more tedious. 

%The appropriate rotation matrix for the spinorial $SO(5)$ representation is 
%\beq
%\Omega(L) = 
%\left [ \ba{cc} \cos ( \ti v/2f) &  i \sin ( \ti v/2f) 
%\\
% i\sin ( \ti v/2f)  &  \cos ( \ti v/2f) 
%\ea \right ]
%\eeq 

%%%%%%%%%%%%%%%%%%%%%%%%%%%%%%%%%%%%%%%%%%%%%%%%%%%%%%%%%%%%%%%%%%%%%%%%%%%%% 
\section{SO(5) generators} 
\setcounter{equation}{0}
\label{a.g} 
%%%%%%%%%%%%%%%%%%%%%%%%%%%%%%%%%%%%%%%%%%%%%%%%%%%%%%%%%%%%%%%%%%%%%%%%%%%%%% 

$SO(5)$ has 10 generators: $T_L^a$ form the $SU(2)_L$ subgroup, 
$T_R^a$ form the $SU(2)_R$ subgroup and the remaining four generators are denoted by $T_C^b$.  
The commutation relations: 
\beq 
\label{e.so5cr}
[T_L^a,T_L^b ] = i \eps^{abc} T_L^c
\qquad  
[T_R^a,T_R^b ] = i \eps^{abc} T_R^c
\qquad 
[T_L^a,T_R^b ] = 0  
\eeq 
\beq 
[T_C^a,T_C^b ] = {i \over 2} \eps^{abc} (T_L^c + T_R^c) 
\qquad  
[T_C^a,T_C^4 ] = {i  \over 2} (T_L^a - T_R^a)
\eeq 
\beq 
[T_{L,R}^a,T_C^b ] = {i \over 2} \left ( \eps^{abc} T_C^c \pm \delta^{ab} T_C^4 \right ) 
\qquad  
[T_{L,R}^a,T_C^4 ] = \mp {i  \over 2} T_C^a
\eeq

The smallest non-trivial $SO(5)$ representation is the spinorial one denoted as ${\bf 4}$.  
The five 4x4 gamma matrices of SO(5):
\beq
\Gamma^a = \left [ \ba{cc} 
0 & \sigma^a 
\\
\sigma^a & 0  
\ea \right ]
\quad 
\Gamma^4 = \left [ \ba{cc} 
0 & - i  
\\
i  & 0  
\ea \right ]
\quad 
\Gamma^a = \left [ \ba{cc} 
1 & 0
\\
0 & -1  
\ea \right ]
\eeq 
The generators: 
\beq
T_L^a = {1 \over 2} \left [ \ba{cc} 
\sigma^a & 0 
\\
0 &  0  
\ea \right ] 
\qquad 
T_R^a = {1 \over 2} \left [ \ba{cc} 
0 & 0 
\\
0 &  \sigma^a 
\ea \right ] 
\eeq 
\beq
T_C^a = {i \over 2 \sqrt{2}} \left [ \ba{cc} 
0 & \sigma^a 
\\
-  \sigma^a  &  0  
\ea \right ] 
\qquad 
T_C^4 = {1 \over 2 \sqrt{2}} \left [ \ba{cc} 
0 & 1 
\\
1 &  0  
\ea \right ] 
\eeq 
They are normalized as $\tr T^\alpha T^\beta = (1/2)\delta^{\alpha\beta}$. 
The $T^3$'s of $SU(2)_L \times SU(2)_R$ are diagonal in this basis.  
Thus, we can easily see how $SU(2)_L \times SU(2)_R$ quantum numbers are embedded in $\bf 4$:
\beq
q = \left [ \ba{c}
q_{+0}
\\
q_{-0} 
\\
q_{0+} 
\\
q_{0-} 
\ea  \right ] 
\qquad 
{\bf 4 = (2,1)\oplus(1,2) }
\eeq  
The Wilson rotation matrix:  
\beq
 \exp(i x T_C^4)  = \left [ \ba{cccc}
  \cos(x/2\rt) &  i \sin(x/2\rt)
  \\ 
  i\sin(x/2\rt) & \cos(x/2\rt)
\ea 
\right ]
\eeq

The fundamental $SO(5)$ representation is denoted as ${\bf 5}$.  
The generators can be chosen as: 
\bea
T_{L,ij}^a & = &  
-{i \over 2} \left [ 
{1 \over 2} \eps^{abc}(\delta_i^b \delta_j^c - \delta_j^b \delta_i^c)
+ (\delta_i^a \delta_j^4 - \delta_j^a \delta_i^4)
\right]  \quad  a = 1 \dots 3
\nn 
T_{R,ij}^a & = & 
-{i \over 2} \left [ 
{1 \over 2} \eps^{abc}(\delta_i^b \delta_j^c - \delta_j^b \delta_i^c)
-(\delta_i^a \delta_j^4 - \delta_j^a \delta_i^4)
 \right]  \quad  a = 1 \dots 3
\nn 
T_{C,ij}^\ha & = &  
-{i \over \sqrt 2} \left [ 
\delta_i^\ha \delta_j^5 - \delta_j^\ha \delta_i^5) 
 \right]  \qquad  \qquad  \ha = 1 \dots 4
\eea 
and are normalized as $\tr T^\alpha T^\beta = \delta^{\alpha\beta}$.

The $T^3$ generators of the $SU(2)_L \times SU(2)_R$ subgroup are non-diagonal in this basis.  rators of the $SU(2)_L \times SU(2)_R$ subgroup are non-diagonal in this basis.  
The vector of $SO(5)$ can be expressed as a combination of eigenvectors of $T^3_L \times T^3_R$,    
\beq
Q = {1 \over \rt} \bvec 
q_{++} + q_{--}
\\
i q_{++} -  i q_{--}
\\
 q_{+-} + q_{-+}
\\
i q_{+-} -  i q_{-+}
\\
\rt q^c
\evec
\eeq   
where $\pm$ denotes $\pm 1/2$. 
Thus, $\bf 5$ contains a bifundamental and a singlet under $SU(2)_L \times SU(2)_R$.  

The Wilson rotation matrix: 
\beq
e^{i x T_C^4} = \left [ \ba{ccccc}
 1& 0& 0& 0& 0\\
 0& 1& 0& 0& 0\\ 
 0& 0& 1& 0& 0\\ 
 0& 0& 0& \cos(x/\rt)& \sin(x/\rt)\\ 
 0& 0& 0& - \sin(x/\rt)& \cos(x/\rt)
\ea 
\right ]
\eeq    
We also show how these matrices operate in the subspace $(q_{+-}, q_{-+},q^c)$: 
\beq
T_C^4 = {1 \over 2} 
\left [ \ba{ccc}
 0& 0 & -1\\
 0& 0 & 1 \\ 
 -1& 1& 0 
\ea 
\right ]
\quad
e^{i h T_C^4} = \left [ \ba{ccc}
{ 1 +  \cos(h/\rt) \over 2} &  {1 -  \cos(h/\rt) \over 2} & - i {\sin(h/\rt)  \over \sqrt 2} \\
  {1 -  \cos(h/\rt) \over 2} &  {1 +  \cos(h/\rt) \over 2} & i {\sin(h/\rt) \over \sqrt 2}  \\ 
-  i {\sin(h/\rt)  \over \sqrt 2} & i{ \sin(h/\rt) \over \sqrt 2} & \cos(h/\rt)
\ea 
\right ]
\eeq

%&&&&&&&&&&&&&&&&&&&&&&&&&&&&&&&&&&&66

\end{document}